\documentclass[
  aps,
  prl,
  reprint,
  superscriptaddress,
  nobibnotes
]{revtex4-2}

\usepackage[utf8]{inputenc}
\usepackage[T1]{fontenc}
\usepackage{lmodern}
\usepackage{graphicx}  
\usepackage{color}
\usepackage{amsmath}
\usepackage{amssymb}
\usepackage{dcolumn}
\usepackage{bm}
\usepackage[colorlinks,bookmarks=true,citecolor=blue,linkcolor=blue,urlcolor=blue]{hyperref}

\begin{document}
\preprint{AIP/123-QED}
\title{Hidden Ergodic Relaxation in the Quench Dynamics of a Bichromatic Mott Lattice}
\author{Áttis~V.~M.~Marino}
\affiliation{Instituto de Física de São Carlos, Universidade de São Paulo, CP 369, 13560-970 São Carlos, SP, Brazil.}
\author{Rhombik~Roy}
\affiliation{Department of Physics, University of Haifa, Haifa 3498838, Israel.}
\affiliation{Haifa Research Center for Theoretical Physics and Astrophysics, University of Haifa, Haifa 3498838, Israel.}
\author{M. A. Caracanhas}
\affiliation{Instituto de Física de São Carlos, Universidade de São Paulo, CP 369, 13560-970 São Carlos, SP, Brazil.}
\author{N.~D.~Chavda}
\affiliation{Department of Applied Physics, Faculty of Technology and Engineering,The Maharaja Sayajirao University of Baroda, Vadodara-390001, India.} 
\author{Antônio~M.~S.~Macêdo}
\affiliation{ Laboratório de Física Teórica e Computacional, Departamento de Física, Universidade Federal de Pernambuco, 50670-901 Recife, Pernambuco, Brazil}
\author{V.~S.~Bagnato}
\affiliation{Instituto de Física de São Carlos, Universidade de São Paulo, CP 369, 13560-970 São Carlos, SP, Brazil.}
\affiliation{Department of Biomedical Engineering, Texas A\&M University, College Station, Texas 77843, USA}
\author{Robin~P~Sagar}
\affiliation{Departamento de Química, Universidad Autónoma Metropolitana -Iztapalapa, Rafael Atlixco 186, CDMX, 09340, Ciudad de México, México.}
\author{Barnali~Chakrabarti}
\email{barnali@if.usp.br}
\affiliation{Instituto de Física de São Carlos, Universidade de São Paulo, CP 369, 13560-970 São Carlos, SP, Brazil.}
\affiliation{ Laboratório de Física Teórica e Computacional, Departamento de Física, Universidade Federal de Pernambuco, 50670-901 Recife, Pernambuco, Brazil}

\date{\today} 

\begin{abstract}

We investigate the nonequilibrium dynamics of strongly interacting bosons in a finite bichromatic Mott lattice following a sudden quench of the secondary lattice amplitude. The coefficient entropies and second R\'enyi entropy exhibit pronounced growth toward their Gaussian Orthogonal Ensemble (GOE) predictions from random-matrix theory, consistent with GOE-like statistical spreading in the employed multiconfigurational representation. In striking contrast, experimentally accessible observables, including the momentum distribution, fragmentation, and Glauber correlation functions, remain nearly unchanged throughout the evolution. For the sampled strong quenches, the coefficient entropies, second R\'enyi entropy, and the $N$-body coefficient spreading collapse onto a common relaxation trajectory that becomes largely independent of the perturbation strength. Our results reveal an emergent hidden ergodic relaxation beneath the persistent local Mott-like order. An effective embedded random-matrix model captures the qualitative crossover from restricted to extensive Hilbert-space spreading, providing an interpretive framework for the observed relaxation dynamics. 

\end{abstract}


\maketitle

\noindent\textit{Introduction.—}
Understanding relaxation and thermalization in isolated quantum many-body systems remains a central challenge in nonequilibrium physics~\cite{Rigol2008,Polkovnikov2011,Eisert2015}. Quantum quenches can produce diverse relaxation phenomena, including prethermalization~\cite{Berges_2004,Langen_2016}, relaxation toward a generalized Gibbs ensemble (GGE)~\cite{Rigol,Langen_2015}, universal scaling dynamics, and non-thermal fixed points~\cite{Pr_fer_2018,Erne_2018}. Ultracold atoms in optical lattices provide a highly controllable platform for exploring such
dynamics owing to their tunable interactions and isolation from the environment~\cite{Bloch2008,Lewenstein2007,Greiner2002}. Strongly interacting bosons
exhibit fragmentation, and correlation-driven relaxation following quantum quenches~\cite{DAlessio2016,Borgonovi2016}.

Entropy measures, participation ratios, and entanglement growth provide sensitive probes of Hilbert-space spreading and ergodic many-body dynamics~\cite{Rigol2008,Eisert2015,DAlessio2016,Borgonovi2016,Kaufman2016,Neill2016,Molignini:2024,Chakrabarti2025PRE}. In sufficiently complex interacting systems, the many-body wavefunction can approach statistical properties described by random-matrix theory (RMT)~\cite{Borgonovi2016,Izrailev_2012,Zelevinsky1996,Srednicki_1994}. An important open question is whether such global many-body relaxation necessarily manifests itself in experimentally accessible low-order observables, particularly in finite strongly correlated systems. Here, we investigate strongly interacting bosons in a finite bichromatic Mott lattice following a sudden quench of the secondary lattice potential. We employ the multiconfigurational time-dependent Hartree method for bosons (MCTDHB), which captures many-body correlations. We find a striking separation between global many-body relaxation and local observables: while the coefficient Shannon entropy and second R\'enyi entropy exhibit extensive Hilbert-space spreading and approach their Gaussian Orthogonal Ensemble (GOE) benchmarks, the momentum distribution, fragmentation, and local two-body Glauber correlations remain close to their initial Mott-like behavior. The short-time fidelity tracks the loss of memory of the initial state, while the Hilbert-space dynamics reveal a clear weak-to-strong-quench crossover, with the sampled strong quenches collapsing onto a common, largely quench-independent relaxation trajectory.

These results reveal a hidden statistical-relaxation regime in which global many-body spreading develops beneath persistent local Mott-like order. We further employ an effective embedded random-matrix model (ERMT) to capture the qualitative crossover from restricted to extensive Hilbert-space spreading. Because MCTDHB uses a time-adaptive basis whereas ERMT uses a fixed many-body basis, their comparison is made at the level of dynamical regimes and effective-shell diagnostics. The basis-independent fidelity provides a direct spectral bridge between the two descriptions through its relation to the local density of states.

\vskip 0.4 cm

\noindent\textit{Model and Protocol.—}
We consider $N=9$ interacting bosons confined in a finite one-dimensional bichromatic optical lattice with $S=9$ lattice sites and hard-wall boundary conditions. The system is described by the many-body Hamiltonian $\hat H = \sum_{i=1}^{N}\left[-\frac{\hbar^2}{2m}\frac{\partial^2}{\partial x_i^2}+V(x_i)\right]+g_0\sum_{i<j}\delta(x_i-x_j)$
where the quasiperiodic lattice potential is $V(x)=V_p\sin^2(k_p x)+V_d\sin^2(k_d x)$,
with irrational ratio $k_d/k_p \simeq 1.197215$. The primary lattice depth is fixed at $V_p=10$, placing the initial state in the strongly correlated Mott regime at unit filling. Non-equilibrium dynamics are initiated by suddenly switching on the secondary lattice $V_d$ at $t=0$ and are governed by the time-dependent many-body Schrödinger equation $\hat{H} \vert\Psi(t) \rangle = i \hbar \frac{\partial}{\partial t}\vert\Psi(t)\rangle$.
\vskip 0.4 cm

\noindent\textit{Methods.—}
The many-body dynamics are computed using MCTDHB implemented in the MCTDH-X package~\cite{Streltsov:2006,Streltsov:2007,Alon:2007,Alon:2008,Lode:2016,Fasshauer:2016,Lode:2020,Lin:2020,MCTDHX}. The many-body wavefunction is expanded in terms of time-dependent permanents; $\vert \Psi(t)\rangle =\sum_{\vec{n}}^{} C_{\vec{n}}(t)\vert \vec{n};t\rangle$ constructed from variationally optimized single-particle orbitals, allowing both the expansion coefficients and orbitals to evolve self-consistently in time.  Imaginary-time propagation is used to obtain the initial ground state, while real-time propagation is employed to study the nonequilibrium quench dynamics. Convergence is verified by increasing the number of orbitals until all relevant observables and entropy measures remain unchanged.
\vskip 0.4 cm

\noindent\textit{Quantities of Interest.—}

To characterize the nonequilibrium dynamics, we compute the momentum distribution $n(k,t)$, the natural occupations $n_i(t)$, the mesoscopic order parameter $\Delta(t)=\sum_i\left(n_i(t)/N\right)^2$, which quantifies fragmentation and coherence loss, and the local two-body Glauber correlation function $g^{(2)}(0,0;t)=\rho^{(2)}(0,0;0,0;t)/[\rho(0,t)]^2$, where $\rho^{(2)}(x_1,x_2;x_1,x_2;t)=\langle\hat{\Psi}^{\dagger}(x_1,t)\hat{\Psi}^{\dagger}(x_2,t)\hat{\Psi}(x_2,t)\hat{\Psi}(x_1,t)\rangle$ and $\rho(x,t)\equiv\rho^{(1)}(x;x;t)$ is the one-body density.

To characterize Hilbert-space spreading, we compute the followings. The coefficient Shannon entropy,
$ S_C(t)=-\sum_{\vec n}|C_{\vec n}(t)|^2\ln|C_{\vec n}(t)|^2$,
quantifies the spreading of the many-body wavefunction over the Fock configuration space.  The second Rényi entropy is defined as $S_2(t)=
-\ln\!\left(
\sum_{\vec n}|C_{\vec n}(t)|^4
\right)$, and quantifies the effective Hilbert-space volume explored by the many-body wavefunction. 

We further introduce the $N$-body coefficient spreading; $S_C^{N}=
-\sum_{\vec n,\vec n'}
\sqrt{p_{\vec n}(t)p_{\vec n'}(t)}
\ln\!\left[
\sqrt{p_{\vec n}(t)p_{\vec n'}(t)}
\right],
$
with $p_{\vec n}(t)=|C_{\vec n}(t)|^2$, incorporates pairwise contributions from all many-body configurations and therefore provides a more global measure of the complexity of the coefficient distribution, but not an entropy in the conventional information-theoretic sense. While the orbital entropy, $S_n(t)=
-\sum_{i=1}^{M}
\frac{n_i(t)}{N}
\ln\!\left(\frac{n_i(t)}{N}\right),
$ quantifies the distribution of bosons among the natural orbitals and therefore provides an information-theoretic measure of fragmentation. Finally, the fidelity, $ F(t)=
|\langle\Psi(0)|\Psi(t)\rangle|^2,
$ measures the overlap of the evolving many-body state with the initial state and quantifies the loss of memory following the secondary-lattice quench. The combined evolution of $S_C$, $S_2$,$S_C^{N}$, $S_n$ and $F$ provides complementary in-
formation on coefficient spreading, fragmentation, and
memory loss.

\vskip 0.4 cm

\noindent\textit{Initial State and Quench Dynamics.—}
We focus on the strongly interacting regime with $g_0=0.5$, where the ground state of the primary lattice at unit filling ($N=S=9$) forms a fully fragmented Mott-insulating state. In this regime, the one-body density is localized with approximately one particle per site and negligible density overlap between neighboring wells. The natural occupations satisfy $n_1 \approx n_2 \approx \cdots \approx n_9 \approx 1/9$, indicating nearly equal occupation of nine orbitals and strong many-body fragmentation beyond a mean-field description. The first-order coherence is suppressed between different lattice sites, while the second-order correlations exhibit pronounced correlation holes, signaling strong suppression of double occupancy and robust local correlations characteristic of a strongly correlated Mott phase. Starting from this correlated initial state, the nonequilibrium dynamics are initiated by suddenly switching on the secondary quasiperiodic lattice potential at $t=0$.

\vskip 0.4 cm
\noindent\textit{Robustness of Local Observables.—}

\begin{figure}[!tbh]
    \centering
    \includegraphics[width=\columnwidth]{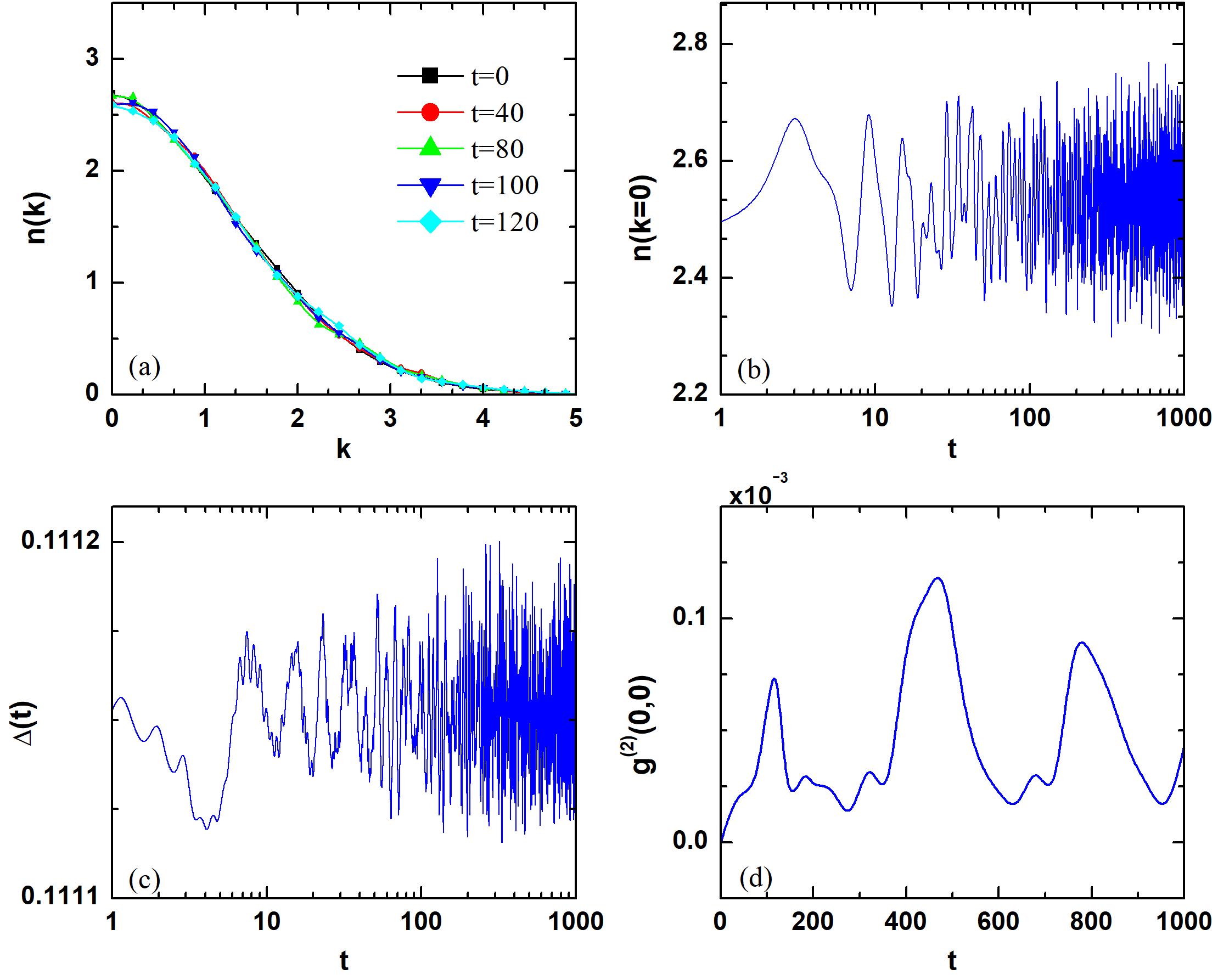}
\caption{Dynamics of experimentally accessible observables following a sudden quench of the secondary quasiperiodic lattice for $N=S=9$. (a) Momentum distribution $n(k,t)$, (b) zero-momentum occupation $n(k\!\to\!0,t)$, (c) order parameter $\Delta(t)$, and (d) local two-body Glauber correlation $g^{(2)}(0,0;t)$. Their weak temporal variations demonstrate the persistence of Mott-like order. Parameters are $V_p=10$, $V_d=5$, $g_0=0.5$, and $M=10$.}
    \label{Fig1}
\end{figure}

We first analyze the nonequilibrium dynamics of experimentally accessible observables following the sudden activation of the secondary quasiperiodic lattice. Fig.~\ref{Fig1} summarizes the evolution of the momentum distribution $n(k,t)$, the zero-momentum occupation $n(k\!\to\!0,t)$, the order parameter $\Delta(t)$, and the local two-body Glauber correlation function $g^{(2)}(0,0;t)$ under sudden secondary-lattice quench.

Fig.~\ref{Fig1}(a) shows the momentum distribution $n(k,t)$ at different evolution times for a strong quench amplitude $V_d=5$. Remarkably, the momentum profiles remain nearly indistinguishable throughout the dynamics, revealing an unexpected robustness of the Mott state against the quench. No substantial redistribution of spectral weight is observed despite the strong nonequilibrium excitation.

The stability of the low-momentum occupation is further illustrated in Fig.~\ref{Fig1}(b), where we plot $n(k\!\to\!0,t)$  on a
logarithmic time scale. After a short initial transient, the zero-momentum occupation exhibits only weak smooth oscillations around its mean value, without any appreciable growth associated with phase coherence or condensate formation.

Fig.~\ref{Fig1}(c) presents the evolution of the order parameter $\Delta(t)$ on a logarithmic time scale. The order parameter fluctuates only weakly around the fragmented Mott value $\Delta \simeq 1/S$, indicating that the system remains strongly fragmented during the entire evolution. The quench therefore does not induce any significant restoration of coherence between lattice sites. All panels are computed with $M=10$ orbitals.

Finally, Fig.~\ref{Fig1}(d) shows the local two-body Glauber correlation function $g^{(2)}(0,0;t)$. The correlation dynamics display only small temporal variations and remain close to their initial strongly correlated Mott values, demonstrating the persistence of local interaction-driven correlations even under strong nonequilibrium driving.

Taken together, the weak dynamical response of $n(k,t)$, $n(k\!\to\!0,t)$, $\Delta(t)$, and $g^{(2)}(0,0;t)$ demonstrates that the quench neither restores phase coherence nor induces substantial delocalization. Instead, the system preserves its localized Mott-like character throughout the evolution. These results reveal an apparent dynamical freezing of low-order observables, suggesting that experimentally accessible quantities may significantly underestimate the complexity of the underlying many-body dynamics.
\vskip 0.4 cm

\noindent\textit{Entropy Growth, Universality, and Hidden Hilbert-Space Relaxation.—}

\begin{figure}[!tbh]
    \centering
    \includegraphics[width=0.9\columnwidth]{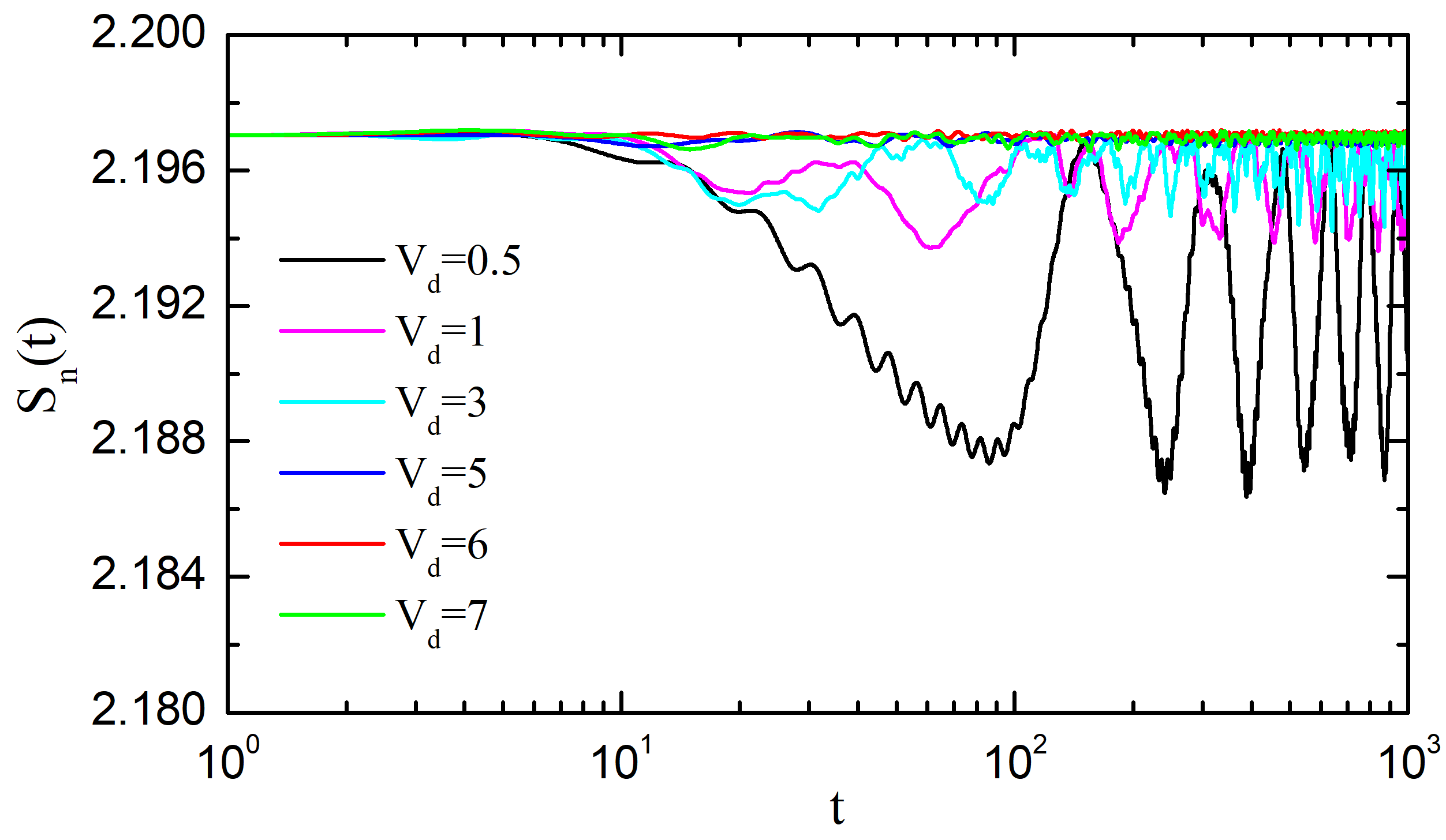}
\caption{
Time evolution of the orbital entropy $S_n(t)$ for different quench amplitudes $V_d$. Strong quenches ($V_d=5,6,7$) rapidly reach a stationary value, while weak quenches retain fluctuations, consistent with persistent Mott-like correlations at the orbital level.}
    \label{Fig2}
\end{figure}

The persistence of Mott-like behavior in low-order observables raises the question of whether the many-body wavefunction nevertheless undergoes substantial reorganization in Hilbert space. Since momentum distributions, and local two-body correlations probe only low-order reduced sectors, they may remain insensitive to the redistribution of amplitudes among many-body configurations. We therefore examine the orbital entropy $S_n(t)$, the coefficient Shannon entropy $S_C(t)$, the second R\'enyi entropy $S_2(t)$, the $N$-body coefficient spreading $S_C^N(t)$, and the fidelity $F(t)$. Together, these quantities provide complementary probes of fragmentation, many-body wavefunction spreading, and loss of memory of the initial state. 
Fig.~\ref{Fig2} shows the orbital entropy $S_n$ for different quenches. For strong quenches ($V_d=5,6,7$), $S_n$ rapidly approaches $\overline{S_n}\approx2.19$, consistent with the mesoscopic order parameter $\Delta\simeq0.11$. Their near-constant values indicate negligible changes in fragmentation and persistence of Mott-like correlations at the orbital level, in contrast to the Hilbert-space spreading revealed below.

\begin{figure}[!tbh]
    \centering
    \includegraphics[width=0.8\columnwidth]{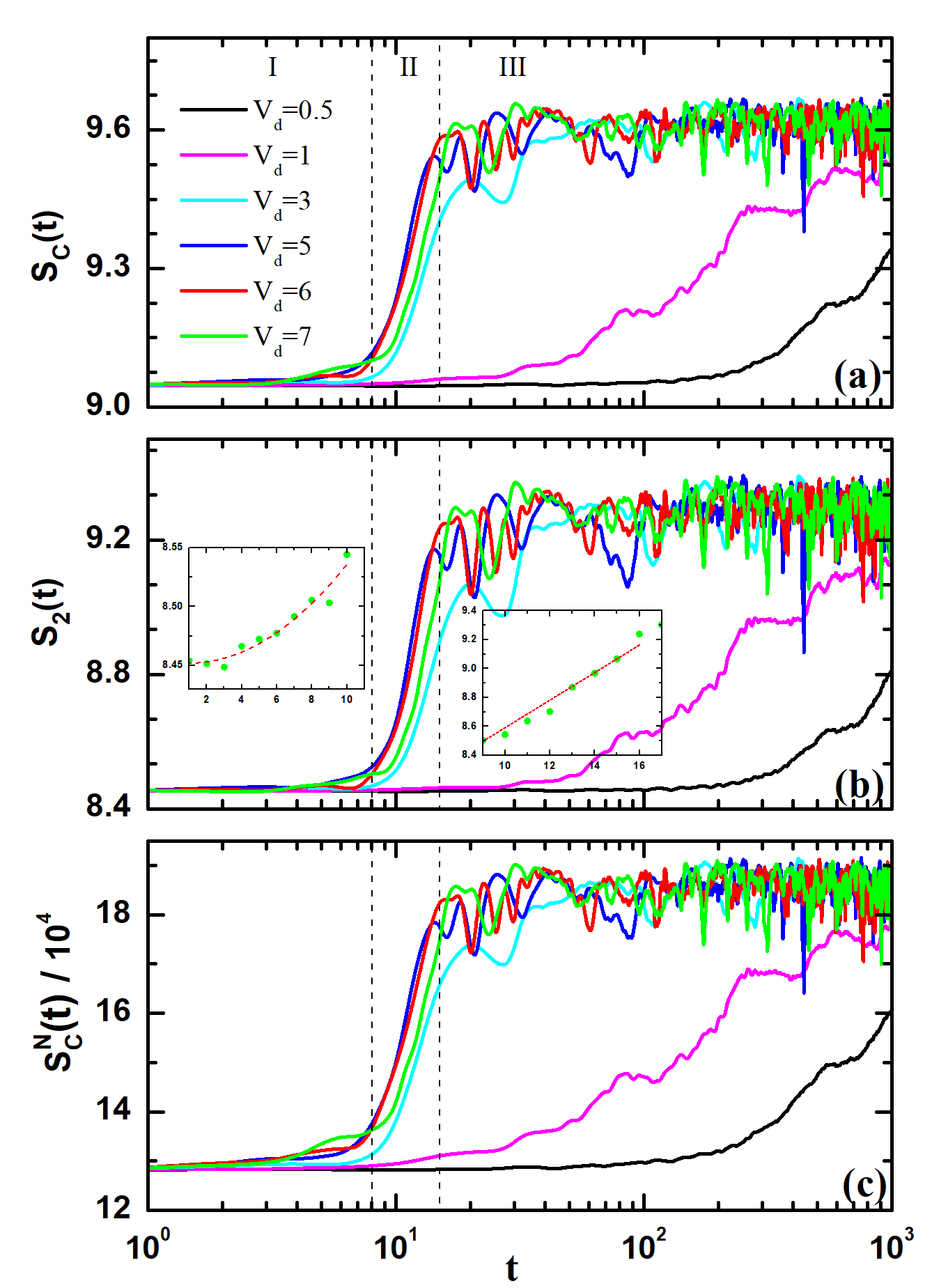}
\caption{Hilbert-space spreading after a sudden quench for $N=S=9$. (a) Coefficient entropy $S_C(t)$, (b) second R\'enyi entropy $S_2(t)$, and (c) $N$-body coefficient spreading $S_C^N(t)$. Weak quenches show restricted spreading, while strong quenches relax toward GOE values. Insets show the initial quadratic and intermediate linear regimes of $S_2(t)$. Parameters: $V_p=10$ and $g_0=0.5$.}
    \label{Fig3}
\end{figure}

Fig.~\ref{Fig3} presents the time evolution of $S_C$, $S_2$, and $S_C^N$ over a broad range of quench amplitudes $V_d$. These measures reveal a clear separation between weak and strong quenches. Weak quenches produce restricted Hilbert-space spreading, whereas strong quenches exhibit a common three-stage relaxation with an initial perturbative onset, an intermediate spreading regime, and long-time saturation with bounded fluctuations.

At short times, all quench amplitudes exhibit the same quadratic onset of transition probabilities, reflecting perturbative unitary dynamics. Expanding the evolution operator, $|\Psi(t)\rangle=e^{-i\hat{H}t}|\Psi(0)\rangle =(1-i\hat{H}t-\hat{H}^{2}t^{2}/2+\cdots)|\Psi(0)\rangle$ an initially unoccupied configuration $|m\rangle$ acquires amplitude $C_m(t)=\langle m|\Psi(t)\rangle\simeq-it\langle m|\hat{H}|\Psi(0)\rangle$, and hence $p_m(t)= |C_m(t)|^2\simeq t^2|\langle m|\hat{H}|\Psi(0)\rangle|^2$

Thus, initially empty configurations acquire population quadratically in time, with the quench dependence entering through the corresponding matrix elements. This perturbative onset gives the quadratic leading behavior of $S_2$ (inset, Zone I) and an apparently quadratic onset of the coefficient measures over the resolved numerical window.

For weak quenches, exemplified by $V_d=0.5$, the subsequent spreading remains slow and the three measures stay far below their eventual strong-quench values, indicating that the wavefunction remains confined to a restricted subset of many-body configurations. At $V_d=1.0$, spreading is enhanced, signaling a crossover toward more efficient many-body mixing, but the available evolution time is still insufficient for complete relaxation. In contrast, for $V_d=5,6,7$, the dynamics rapidly leave the perturbative regime and enter an extended interval of efficient many-body configuration mixing. The approximately linear growth observed in this intermediate time window (inset, Zone II) reflects the progressive occupation of an increasing number of many-body configurations. This behavior is a numerical signature of configuration mixing rather than a universal analytical law.

The three measures therefore display the same qualitative sequence of early growth, intermediate spreading, and long-time saturation, although their strict short-time asymptotics are not identical. If the initial state occupies a single Fock configuration and initially empty configurations acquire probabilities $p_m(t)=\gamma_m t^2+\cdots, $ then
\[
\begin{aligned}
\text{Zone I:}\qquad
S_C(t) &\sim 2\kappa t^2\ln\frac{1}{|t|}+O(t^2),\\
S_2(t) &\sim 2\kappa t^2+O(t^4),\\
S_C^N(t) &\sim 2B|t|\ln\frac{1}{|t|}+O(|t|),\\[1mm]
\text{Zone II:}\qquad
S_C(t) &\sim a_C+\Gamma_C t,\\
S_2(t) &\sim a_2+\Gamma_2 t,\\
S_C^N(t) &\sim a_N+\Gamma_N t,\\[1mm]
\text{Zone III:}\qquad
S_C(t) &\sim S_C^{\mathrm{sat}},\\
S_2(t) &\sim S_2^{\mathrm{sat}},\\
S_C^N(t) &\sim S_C^{N,\mathrm{sat}},
\end{aligned}
\]
where  $\kappa=\sum_{m\neq 0}\gamma_m,
\qquad B=\sum_{m\neq 0}\sqrt{\gamma_m}.
$

Thus, the apparently quadratic Zone-I growth in the resolved numerical window is only an empirical characterization: strictly as $t\to0$, $S_C$ contains a $t^2\ln(1/|t|)$ contribution, while $S_C^N$ can begin as $|t|\ln(1/|t|)$; only $S_2$ has a genuinely analytic quadratic leading term.

Remarkably, the strong-quench curves for $V_d=5,6,7$ collapse onto a common relaxation trajectory. This indicates that, within the sampled strong-quench window, the spreading dynamics become governed predominantly by the many-body Hilbert-space structure rather than by the microscopic quench amplitude. The resulting quench-independent behavior provides evidence for an emergent universal relaxation regime.

At long times, the three measures saturate and fluctuate around well-defined mean values, signaling relaxation toward a statistically stationary state. The saturation values can be compared with random-matrix theory for the Gaussian Orthogonal Ensemble (GOE)~\cite{Borgonovi2016,Izrailev_2012,Zelevinsky1996}. For a Hilbert-space dimension $D$, $S_C^{\mathrm{GOE}}=\ln(0.48D)$ and $S_2^{\mathrm{GOE}}=\ln(D/3)$.

Taking the nominal dynamically relevant orbital number $M_{\mathrm{eff}}=9$ gives $D=\binom{N+M_{\mathrm{eff}}-1}{N}=24310$. Consequently, the GOE estimates are $S_C^{\mathrm{GOE}}\approx\ln(0.48D)\approx 9.36$ and $S_2^{\mathrm{GOE}}\approx\ln(D/3)\approx 8.99$. The observed plateaus,$\overline{S_C}\approx9.6,\qquad \overline{S_2}\approx9.23,$ are in close agreement with these GOE values, while $S_C^N$ independently approaches a stationary value with small fluctuations. This agreement supports GOE-like statistical spreading of the many-body wavefunction in the accessible Hilbert-space sector.

Because the MCTDHB basis is time adaptive, the nominal $M_{\mathrm{eff}}=9$ dimension should be regarded as a reference rather than as a strict identification of the dynamically explored shell. An internal estimate based on the two independent coefficient moments gives 
$ D_{\mathrm{eff}}^{(S_C)} =\exp(\overline{S_C})/0.48, \qquad D_{\mathrm{eff}}^{(S_2)} =3\exp(\overline{S_2})$ yielding $D_{\mathrm{eff}}^{(S_C)}\simeq3.08\times10^4, \qquad D_{\mathrm{eff}}^{(S_2)}\simeq3.06\times10^4.$
Their close agreement provides an internal consistency check on the GOE-like coefficient statistics and indicates that the dynamics explore a substantial, but not complete, portion of the full $M=10$ combinatorial space, $D=48620$. Establishing basis-independent many-body ergodicity would require a fixed-basis projection or complementary spectral diagnostics.

Finally, the remaining bounded fluctuations are expected for an isolated finite many-body system and reflect the finite accessible Hilbert-space sector. Thus, the strong-quench dynamics reveal extensive statistical wavefunction spreading even though the orbital observables retain the Mott-like structure identified in Fig.~\ref{Fig2}.

\begin{figure}[!tbh]
    \centering
    \includegraphics[width=0.8\columnwidth]{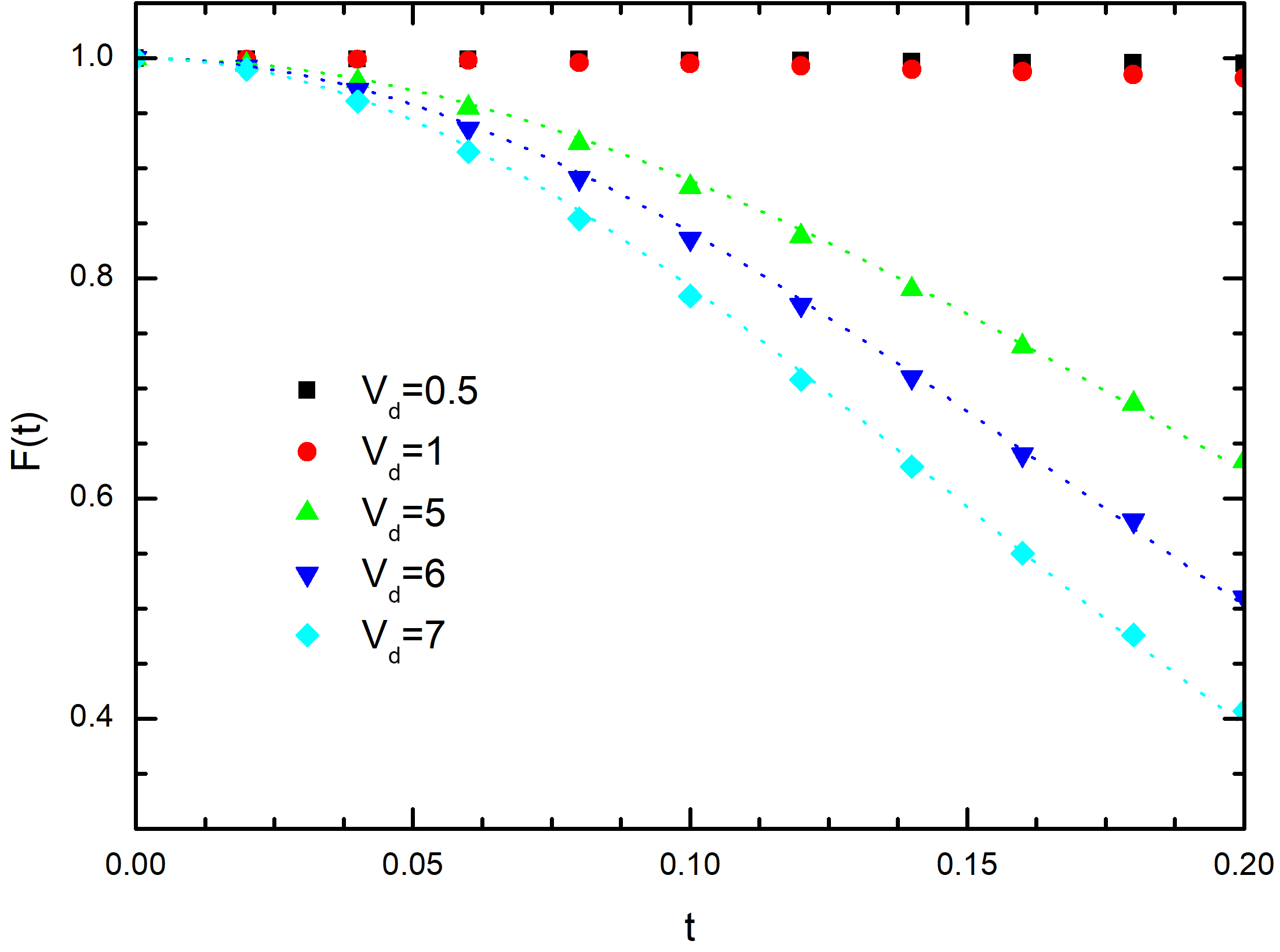}
\caption{Short-time fidelity $F(t)$ after sudden quenches of the secondary quasiperiodic lattice for different quench amplitudes $V_d$ with $N=S=9$. The faster decay for stronger quenches reflects enhanced many-body dephasing and rapid loss of memory of the initial state. Parameters are $V_p=10$, $g_0=0.5$.}
    \label{Fig4}
\end{figure}

The growth of Hilbert-space complexity does not directly quantify the loss of  memory of the initial state. To characterize this complementary aspect of the dynamics, we analyze the fidelity  $F(t)=|\langle\Psi(0)|\Psi(t)\rangle|^2$, which measures the overlap between the evolving many-body state and the initial Mott state. Fig.~\ref{Fig4} presents the short-time fidelity dynamics for different quench amplitudes $V_d$. For weak quenches, $F(t)$ remains close to unity over the time window shown, indicating that the wavefunction remains confined near the initial configuration and that Hilbert-space spreading is strongly suppressed. As the quench strength increases, the fidelity exhibits a faster decay, reflecting enhanced many-body mixing and progressive loss of memory of the initial state. 

At short times, the fidelity follows the universal quadratic decay expected from unitary perturbation theory, $F(t)=1-\sigma_E^2 t^2+\mathcal{O}(t^4),$ where $\sigma_E^2=\langle\Psi(0)|\hat{H}^2|\Psi(0)\rangle-\langle\Psi(0)|\hat{H}|\Psi(0)\rangle^2$ is the energy variance of the initial state with respect to the post-quench Hamiltonian. The coefficient $\sigma_E$ depends on the quench amplitude $V_d$, since the post-quench Hamiltonian and the associated coupling between many-body configurations change with the perturbation strength. Consequently, stronger quenches produce a larger energy variance and a faster initial decay of the fidelity. While this short-time decay rate retains information about the microscopic details of the quench, the fidelity curves need not collapse, because their energy variances remain quench dependent. Thus, the fidelity complements the entropy measures by providing a basis-independent probe of the quench-dependent short-time regime, while the common strong-quench trajectory is identified from the coefficient measures.

\noindent\textit{Discussion.-}

Our results demonstrate that an isolated many-body system can undergo substantial global wavefunction relaxation without visible thermalization in low-order observables. Interaction-induced many-body dephasing spreads probability amplitudes across the dense network of configurations connected to the strongly correlated Mott state, producing extensive Hilbert-space reorganization while preserving local Mott-like correlations. The combined entropy and fidelity dynamics reveal a crossover from quench-dependent short-time evolution to a common strong-quench relaxation regime. While the initial fidelity decay retains information about the microscopic quench through the energy variance, the long-time Hilbert-space measures become largely quench independent and approach GOE-like statistical behavior, indicating that the late-time dynamics are governed increasingly by the intrinsic structure of the accessible many-body space.

The observed hidden statistical relaxation provides an intrinsic route toward stationarity in an isolated quantum system, without external dissipation. Importantly, our evidence concerns wavefunction statistics and should not be interpreted as a complete demonstration of eigenstate ergodicity or thermalization, since no level-spacing, adjacent-gap, or fixed-basis Porter--Thomas analysis is performed. The persistence of low-order observables despite extensive many-body information spreading further suggests that entropy-based diagnostics can reveal relaxation hidden from conventional measurements, with possible relevance to complex quantum fluids and turbulent Bose gases~\cite{Garc_a_Orozco_2022,Moreno_Armijos_2025}.

\vskip 0.2 cm
B.C. gratefully acknowledges Thomas Gasenzer for insightful discussions and guidance.
This work was supported by FAPESP Grants 2013/07276-1, 2024/04637-8, and
2025/00547-7; CNPq Grant No.~386392/2024-2; and CAPES Grant No.~
88887.252975/2026-00.

\bibliography{manuscript}

\clearpage
\onecolumngrid

\section*{Supplemental Material}

\section{Effective Embedded Random-Matrix Description of Hidden Hilbert-Space Relaxation}

\subsection{Effective embedded random-matrix model (ERMT)}

The MCTDHB simulations show that a sudden secondary-lattice quench can induce extensive spreading of the many-body wavefunction in Hilbert space, while experimentally accessible low-order observables remain close to their initial Mott-like behavior. To identify the generic mechanism responsible for this hidden relaxation, we introduce an effective embedded random-matrix description of the post-quench dynamics. The purpose of this model is not to reproduce the microscopic bichromatic lattice Hamiltonian quantitatively, but rather to capture the generic mechanism of many-body configuration mixing generated by few-body interactions.

The effective Hamiltonian is written as $\hat H_{\mathrm{eff}}=\hat H_0+\lambda(V_d)\hat V$, where $\hat H_0$ represents the unperturbed many-body energy structure and $\hat V$ is an embedded random interaction containing one- and two-body matrix elements~\cite{Kota_2001,Haldar_2016}. The dependence on the quench amplitude $V_d$ is incorporated through the effective coupling strength $\lambda(V_d)$, which controls the degree of configuration mixing. Unlike a full random matrix, the embedded interaction preserves the few-body character of the microscopic Hamiltonian while allowing complex many-body dynamics to emerge in the Fock basis. The map $\lambda(V_d)$ is phenomenological: it classifies weak, intermediate, and saturated strong mixing but is not derived from the microscopic MCTDHB Hamiltonian. Accordingly, equal labels $V_d$ in the two calculations denote corresponding dynamical regimes, not a calibrated equality of matrix elements or time scales.

A further distinction is essential. The ERMT coefficients refer to a fixed Fock basis, whereas both the orbitals and permanents used by MCTDHB evolve in time. Direct equality of coefficient entropies across the two descriptions is therefore neither assumed nor tested. The comparison concerns the ordering of dynamical regimes, the saturation of the effective shell, and basis-independent survival dynamics.

Expanding the many-body wavefunction in the Fock basis, $|\Psi(t)\rangle=\sum_{\vec n} C_{\vec n}(t)|\vec n\rangle$, the interaction term generates couplings between many-body configurations, $ V_{\vec n\vec m}=\langle\vec n|\hat V|\vec m\rangle$, where the connected configurations differ by the allowed one- and two-body processes of the embedded ensemble. Increasing the effective coupling enhances the mixing between these configurations. In the weak-coupling regime, the wavefunction remains localized over a restricted set of Fock states, resulting in limited Hilbert-space spreading. In contrast, strong coupling produces extensive configuration hybridization and drives the many-body state toward a statistically complex distribution over the accessible Hilbert space.

This provides a minimal framework for interpreting the hidden statistical relaxation identified in the MCTDHB simulations: the many-body wavefunction can undergo strong Hilbert-space reorganization even when low-order observables retain signatures of the initial Mott-like state.

\subsection{Local density of states and fidelity decay}

The configuration mixing generated by the embedded interaction can also be viewed in the energy representation through the local density of states (LDOS). As the quench-induced coupling hybridizes many-body configurations, the initial Mott state is distributed over an increasing number of eigenstates of the post-quench Hamiltonian. This distribution is characterized by the LDOS, or strength function, $\rho_{\mathrm{LDOS}}(E)=\sum_{\alpha}|\langle\alpha|\Psi(0)\rangle|^2 \delta(E-E_\alpha)$, where $|\alpha\rangle$are the eigenstates of $\hat H_{\mathrm{eff}}$. 

A weak quench produces limited configuration mixing and therefore a narrow LDOS, indicating that the initial state remains close to a small number of post-quench eigenstates. In contrast, stronger coupling leads to broader LDOS distributions, reflecting enhanced hybridization among many-body configurations. This provides the spectral interpretation of the increasing Hilbert-space spreading observed in the entropy measures.

The connection between the LDOS and the fidelity follows from the survival amplitude $A(t)=\langle\Psi(0)|e^{-i\hat H_{\mathrm{eff}}t}|\Psi(0)\rangle =\int dE\,\rho_{\mathrm{LDOS}}(E)e^{-iEt}$, with the fidelity given by $F(t)=|A(t)|^2$. Thus, the short-time fidelity decay probes the energy spread generated by the quench, whereas the entropy measures characterize the corresponding redistribution of probability amplitudes in Hilbert space.

In this way, the LDOS provides the spectral link between the ERMT description and the MCTDHB dynamics: although the LDOS is not reconstructed explicitly from the present time-adaptive MCTDHB calculation, its Fourier-transform consequences are tested by the basis-independent fidelity; stronger quench-induced mixing broadens the energy distribution of the initial state, producing faster dephasing and more extensive Hilbert-space spreading.

\subsection{Hilbert-space spreading and GOE relaxation}
The broadening of the LDOS provides a spectral signature of the increasing number of many-body eigenstates participating in the dynamics. In the Fock basis, this process appears as a progressive redistribution of probability amplitudes among many-body configurations, which is quantified by the entropy measures introduced in the main text. Therefore, the growth of these quantities reflects the same underlying mechanism as the LDOS broadening: the interaction-induced hybridization of many-body configurations.

For weak quenches, the coupling between configurations is insufficient to produce extensive mixing, and the wavefunction remains localized within a limited region of Hilbert space. Consequently, the entropy measures exhibit slow evolution and retain a strong dependence on the details of the quench. As the effective coupling increases, more configurations become hybridized and the wavefunction explores a progressively larger fraction of the accessible Hilbert space. In this regime, the relaxation becomes increasingly governed by the statistical properties of the many-body spectrum rather than by the microscopic details of the perturbation.

For sufficiently strong coupling, the accessible Hilbert-space region is expected to become statistically populated, and the entropy measures should approach their Gaussian Orthogonal Ensemble (GOE) expectations. Such GOE-like behavior would support the hypothesis that the embedded random-matrix model captures the essential mechanism responsible for the hidden statistical relaxation observed in the MCTDHB simulations. Full ERMT validation additionally requires consistency between effective dimensions inferred from different coefficient moments and, ideally, fixed-basis Porter–Thomas or spectral diagnostics. Importantly, this statistical spreading occurs at the level of the many-body wavefunction, while low-order observables may still retain signatures of the initial Mott-like state.

Thus, the ERMT framework provides a theoretical connection between quench-induced configuration mixing and the expected relaxation dynamics: weak quenches are expected to produce restricted Hilbert-space spreading, whereas sufficiently strong quenches should lead to extensive many-body reorganization within the accessible Hilbert space. This provides an effective statistical interpretation for the crossover from perturbative dynamics to common strong-quench statistical relaxation that will be tested through the numerical validation below.

\subsection{Numerical Validation of ERMT}
\begin{figure}[t]
    \centering
\includegraphics[width=0.3\columnwidth]{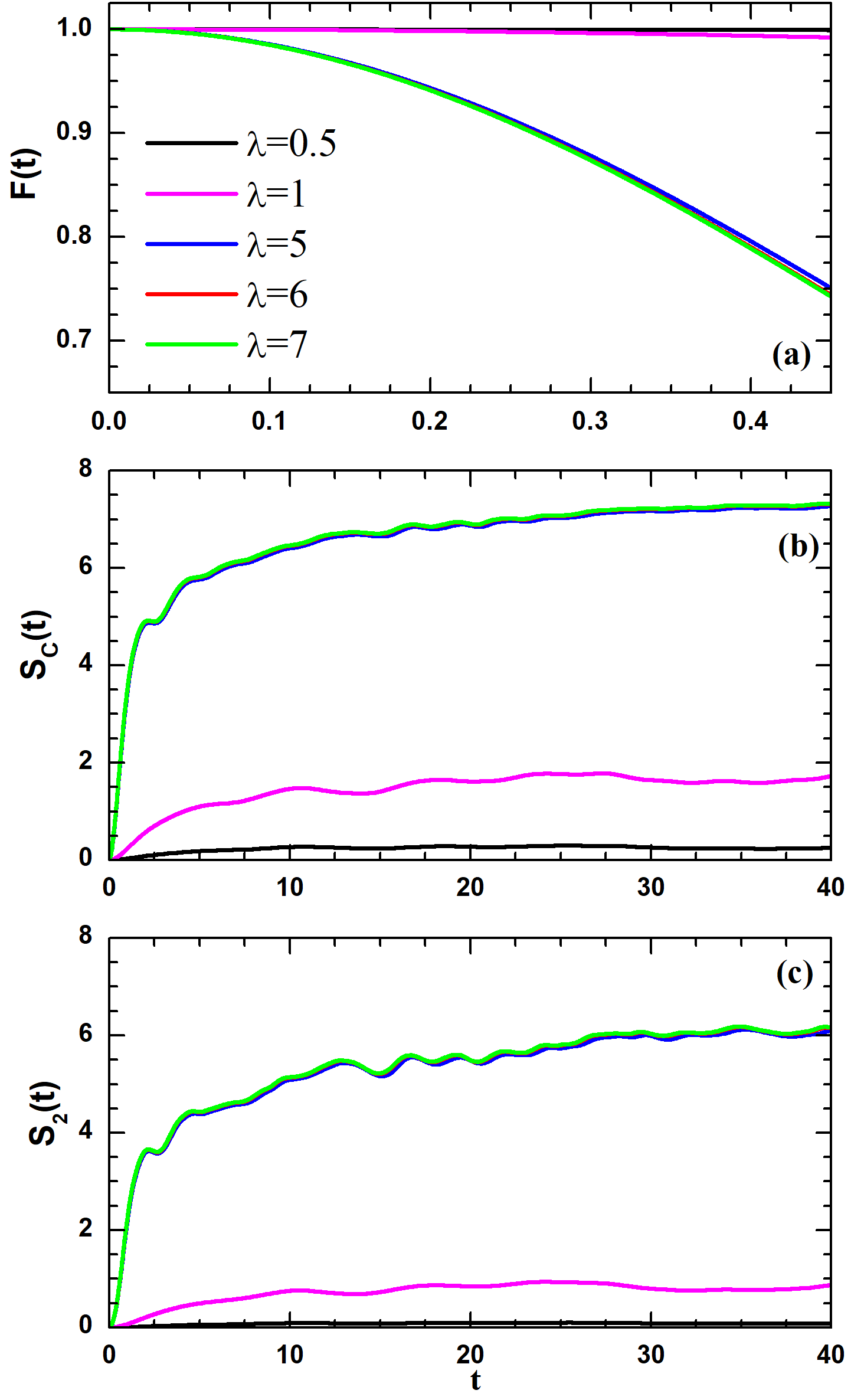}
\caption{Numerical validation of the effective embedded random-matrix model. Dynamics of (a) the fidelity $F(t)$, (b) the coefficient Shannon entropy $S_C(t)$, and (c) the second R\'enyi entropy $S_2(t)$ for different coupling strengths. Weak couplings exhibit limited dephasing and restricted Hilbert-space spreading, whereas strong couplings ($\lambda=5,6,7$) show rapid entropy growth followed by saturation, indicating extensive configuration mixing and a stationary coefficient distribution.}
    \label{fig:ERMT}
\end{figure}

To test whether the effective embedded random-matrix description captures the essential relaxation mechanisms observed in the MCTDHB simulations, we perform independent numerical simulations of the effective Hamiltonian introduced above. The surrogate calculation is performed in a fixed bosonic Fock basis and is deliberately smaller than the microscopic problem; therefore its time axis and saturation values are not compared directly with those of the $N = 9$, $M = 10$, MCTDHB calculation. The purpose of this analysis is not to reproduce the microscopic bichromatic lattice dynamics, but to verify whether the generic consequences of interaction-induced many-body configuration mixing—short-time dephasing, Hilbert-space spreading, and statistical relaxation—emerge in a model with the same few-body structure.

The dynamics are characterized through the fidelity, coefficient Shannon entropy, and second R\'enyi entropy, which probe complementary aspects of the relaxation process. The results are summarized in Fig.~\ref{fig:ERMT}, where the three quantities are shown for different coupling strengths.

Figure~\ref{fig:ERMT}(a) presents the fidelity dynamics for different coupling strengths in the embedded random-matrix model. For weak couplings, the fidelity remains close to unity over the accessible time window, indicating that the initial state undergoes only weak dephasing and limited spreading. This behavior is consistent with the MCTDHB results, where weak quenches preserve a large overlap with the initial Mott state. This agreement is qualitative because the ERMT coupling scale has not been microscopically calibrated.

For all couplings, the ERMT dynamics exhibit the universal short-time quadratic decay expected for unitary evolution, with a coupling-dependent energy variance. The same qualitative behavior is observed in the MCTDHB simulations for large quench amplitudes. In the present manuscript this comparison tests only the quadratic short-time law; neither a Breit-Wigner fidelity window nor the long-time survival plateau is extracted from the microscopic data. However, unlike the ERMT model, where the coupling parameter controls the overall strength of mixing, the microscopic lattice quench generates a $V_d$-dependent energy variance of the initial state. Consequently, the MCTDHB fidelity curves retain distinct decay rates for different strong quenches. The ERMT model captures the universal dephasing mechanism associated with strong configuration mixing, while the detailed dependence of the short-time decay on the quench amplitude remains specific to the microscopic Hamiltonian.

The weak dependence of the ERMT fidelity curves in the strong-coupling regime also reflects the phenomenological saturation imposed through $\lambda(V_d)$. Consequently, the ERMT strong-quench collapse is a consistency check of the saturated-mixing mechanism, not an independent derivation of the microscopic $V_d$ threshold. Once the coupling strength is sufficient to induce extensive configuration mixing, further increases of the coupling do not introduce additional microscopic structure, and the fidelity decay becomes governed primarily by the statistical properties of the ensemble. In contrast, the microscopic MCTDHB Hamiltonian retains a detailed dependence on the quench amplitude through the modified many-body spectrum and coupling structure, resulting in distinct short-time decay rates for different secondary-lattice quenches.

The corresponding growth of the Hilbert-space measures is shown in Fig.~\ref{fig:ERMT}(b,c). The entropy dynamics clearly separate into two distinct regimes. For weak couplings, both \(S_C(t)\) and \(S_2(t)\) exhibit a slow increase over the accessible evolution time, indicating restricted spreading of the many-body wavefunction without reaching a relaxed regime. In contrast, for strong couplings, corresponding to $V_d = 5, 6, 7$ under the chosen saturating map $\lambda(V_d)$, both entropy measures display a rapid increase followed by saturation. In the spreading regime, the coefficient entropy exhibits an approximately linear growth, \(S_C(t)\approx a+\Gamma t\), reflecting the progressive redistribution of probability amplitudes among many-body configurations.

While the ERMT model captures the essential crossover from restricted Hilbert-space spreading to statistical relaxation, it does not reproduce all microscopic features of the MCTDHB dynamics. In particular, the richer three-stage relaxation observed in the bichromatic lattice system originates from the detailed structure of the many-body spectrum and the energy-dependent configuration couplings, which are not explicitly contained in the effective random-matrix description. Furthermore, the saturation values obtained from the ERMT model exhibit deviations from both the GOE expectations and the MCTDHB results. The two calculations therefore test complementary aspects of the dynamics: MCTDHB provides the microscopic many-body evolution, whereas ERMT tests whether sparse connectivity through one- and two-body couplings is sufficient to reproduce the observed sequence of dynamical regimes. Nevertheless, the ERMT results show that interaction-induced configuration mixing is sufficient to generate the hidden Hilbert-space relaxation observed in the microscopic dynamics. 

\subsection{Entropy-based characterization of Hilbert-space spreading}
The saturation values of the entropy measures provide an estimate of the number of many-body configurations effectively involved in the relaxed state. Since the embedded random-matrix simulations reach a stationary plateau without significant long-time fluctuations, the asymptotic values are obtained directly from the numerical curves. Using the GOE-corrected relation for the coefficient entropy, we define an entropy-based effective dimension as

\[
D_{\mathrm{eff}}^{(S_C)}
=
\frac{\exp(S_C^{\mathrm{sat}})}{0.48},
\]

while the second Rényi entropy provides the corresponding raw participation dimension

\[
D_{\mathrm{eff}}^{(S_2)}
=
3\exp(S_2^{\mathrm{sat}}).
\]
For the strong-quench ERMT dynamics, the saturation values
$S_C^{\mathrm{sat}}\approx 7.4$ and $S_2^{\mathrm{sat}}\approx 6.1$ yield

\[
D_{\mathrm{eff}}^{(S_C)}
\approx
\frac{\exp(7.4)}{0.48}
\approx
3.38\times10^{3},
\quad
D_{\mathrm{eff}}^{(S_2)}
\approx
3\exp(6.1)
\approx
1.34\times10^{3}.
\]

The difference between these two estimates reflects the different sensitivity of the entropy measures to the distribution of many-body configuration weights. The coefficient entropy is sensitive to the broad distribution of weakly occupied configurations, whereas the second Rényi entropy is dominated by the configurations carrying the largest probabilities. Therefore, the large mismatch between these two raw values demonstrates extensive mixing but does not, by itself, validate a GOE shell. The appropriate GOE test is agreement between $\exp(S_C^{\mathrm{sat}})/0.48$ and 3 $\exp(S_2^{sat})$, followed by comparison with an independently determined accessible-shell dimension.

These entropy-based effective dimensions should not be interpreted as the fraction of the complete combinatorial Hilbert-space dimension $D=24310$. Instead, they characterize different aspects of the dynamically accessible sector explored after the quench. The fact that both values are much larger than the initial contribution of a single configuration demonstrates substantial many-body configuration mixing and Hilbert-space spreading. Differences from the MCTDHB values are expected, since the ERMT description does not retain the detailed energy-shell structure and configuration-dependent couplings of the microscopic Hamiltonian.

\section{MCTDHB Method}

We employ the multiconfigurational time-dependent Hartree method for bosons (MCTDHB), implemented in the \texttt{MCTDH-X} package~\cite{Streltsov:2006,Streltsov:2007,Alon:2007,Alon:2008,Lode:2016,Fasshauer:2016,Lode:2020,Lin:2020,MCTDHX}. The method solves the many-body time-dependent Schr\"odinger equation using a variationally optimized time-dependent basis, enabling the treatment of correlations beyond mean-field theory.

The many-body wave function is expanded as
\begin{equation}
|\Psi(t)\rangle = \sum_{\vec{n}} C_{\vec{n}}(t)\,|\vec{n};t\rangle,
\end{equation}
where \(\vec{n}=(n_1,\dots,n_M)\) with \(\sum_{k=1}^{M} n_k=N\), and \(M\) denotes the number of time-dependent single-particle orbitals.

The permanents are constructed as
\begin{equation}
|\vec{n};t\rangle =
\prod_{k=1}^{M}\frac{\left(\hat{b}_k^\dagger(t)\right)^{n_k}}{\sqrt{n_k!}}\,|0\rangle,
\end{equation}
with time-dependent creation operators
\begin{equation}
\hat{b}_k^\dagger(t)=\int dx\,\phi_k(x,t)\,\hat{\Psi}^\dagger(x).
\end{equation}
Here, the time-dependent orbitals $\phi_k(x,t)$ form an orthonormal set satisfying $\langle \phi_j(t)|\phi_k(t)\rangle=\delta_{jk}$.

Both coefficients \(C_{\vec{n}}(t)\) and orbitals \(\phi_k(x,t)\) are calculated self-consistently through the coupled equations obtained from the Dirac--Frenkel variational principle. This allows for imaginary-time propagation (ground state) and real-time evolution (nonequilibrium dynamics) within the same framework.

The MCTDHB method systematically converges with increasing number of orbitals $M$. For \(M=1\), the many-body ansatz reduces to a single-orbital mean-field description, corresponding to the Gross-Pitaevskii theory.  In the limit where the orbital set spans the complete single-particle Hilbert space, the MCTDHB expansion becomes formally exact. In practice, numerical convergence is established for finite $M$ by verifying that the relevant observables become insensitive to a further increase in the number of orbitals. In this work, we use \(M=10\) orbitals for \(N=9\) bosons and check convergence by comparison with calculations using different orbital truncations.

The time-dependent natural-orbital occupations provide a direct characterization of the condensation and fragmentation properties of the many-body state. They are obtained by diagonalizing the one-body reduced density matrix,
$\rho^{(1)}(x,x';t) = \langle\Psi(t)|
\hat\Psi^\dagger(x')
\hat\Psi(x)
|\Psi(t)\rangle ,
$ 
which admits the spectral decomposition
$
\rho^{(1)}(x,x';t)=
\sum_i \rho_i(t)
\left(\varphi_i^{(NO)}(x,t)\right) \left(\varphi_i^{(NO)}(x',t)\right)^{*},
$
where the eigenvectors $\varphi_i^{(NO)}(x)$ are the natural orbitals and $\rho_i(t)$ are their corresponding occupations. The occupations satisfy $\sum_i \rho_i(t)=N$, and the normalized natural occupations are defined as $n_i(t)=\frac{\rho_i(t)}{N}$, such that $\sum_i n_i(t)=1$.

The distribution of these natural-orbital occupations provides a direct measure of many-body fragmentation: a single dominant occupation corresponds to a condensed state, whereas several macroscopically occupied natural orbitals indicate fragmentation. In the present work, the time evolution of $n_i(t)$ is used to characterize the persistence of Mott-like correlations at the one-body level and to distinguish orbital-order dynamics from the spreading of the many-body wavefunction over configurations in Hilbert space.

\section{SYSTEM PARAMETERS}

Unless otherwise specified, the simulations presented in the main text are performed for \(N=9\) strongly interacting bosons confined in a finite lattice with \(S=9\) sites and described using \(M=10\) orbitals within the MCTDHB framework. The system is subjected to a bichromatic quasi-periodic optical lattice formed by superimposing a primary lattice potential of depth \(V_p\) and wavelength \(\lambda_p\) with a secondary lattice of depth \(V_d\), wavelength \(\lambda_d\), and relative phase \(\phi\). The resulting external potential is given by
\begin{equation}
V(x)=V_p\sin^2(k_p x)+V_d\sin^2(k_d x+\phi),
\label{S1}
\end{equation}
where the corresponding wave vectors are defined as \(k_i = 2\pi/\lambda_i\). The primary lattice depth is fixed at \(V_p = 10\), while the secondary lattice depth is varied within the range \(V_d \in [0.5,\,7.0]\) in order to investigate the out-of-equilibrium statistical relaxation dynamics. The relative phase is chosen as \(\phi = 0.0\). No phase averaging is performed; consequently, the
results refer to this fixed registration of the two lattices and the hard-wall boundaries.

\subsection*{A. Length Scales}
The system consists of nine lattice sites symmetrically distributed around the origin, corresponding to the spatial interval $x \in [-\frac{9}{2}d,\,\frac{9}{2}d]$, where $d=\pi/k_p=\lambda_p/2$ denotes the primary lattice spacing. The wavelengths of the primary and secondary lattices are chosen as $\lambda_p \approx 532.2~\mathrm{nm}$ and $\lambda_d \approx 444.5~\mathrm{nm}$, respectively, giving an approximately incommensurate ratio $k_d/k_p \approx 1.197215$. The corresponding wave vectors are $k_p \approx 1.1806 \times 10^{7}~\mathrm{m}^{-1}$ and $k_d \approx 1.4135 \times 10^{7}~\mathrm{m}^{-1}$. 
In the numerical implementation, the spatial region is discretized using $512$ grid points, while the characteristic length unit is chosen as $\bar{L}=1/k_p\approx84.7~\mathrm{nm}$. Since the total length of the numerical domain is $9d=9\pi\bar{L}$, the corresponding spatial resolution is approximately
\begin{equation}
\Delta x=\frac{9d}{512}
=\frac{9\pi\bar{L}}{512}
\approx4.68~\mathrm{nm}.
\end{equation}
The corresponding total size of the nine-site system is approximately $2.4  \mu m$.

\subsection*{B. Energy Scales}
For the present work, we consider non-dipolar bosons corresponding to \(^{87}\mathrm{Rb}\) atoms with mass \(m \approx 1.443\times10^{-25}\,\mathrm{kg}\). The unit of energy is defined in terms of the recoil energy associated with the primary lattice,
\begin{equation}
E_r \equiv \frac{\hbar^2 k_p^2}{2m},
\end{equation}
where \(k_p\) denotes the wave vector of the primary lattice. For the chosen lattice parameters, the recoil energy is \(E_r \approx 5.37\times10^{-30}\,\mathrm{J}\), corresponding to the recoil frequency
\begin{equation}
\nu_r=\frac{E_r}{h}\approx8.11~\mathrm{kHz}.
\end{equation}

Throughout the simulations, we employ the characteristic energy unit
\begin{equation}
\bar{E}\equiv\frac{\hbar^2}{m\bar{L}^2}
=2E_r,
\end{equation}
where $\bar{L}=1/k_p\approx84.7~\mathrm{nm}$ represents the characteristic length unit used to nondimensionalize the spatial coordinate. Unless otherwise stated, all energies presented in this work are expressed in units of $E_r$.

The primary lattice depth is fixed at
\begin{equation}
V_p = 10\bar{E}=20E_r,
\end{equation}
while the secondary lattice depth is varied within the interval
\begin{equation}
V_d \in [0.5,7.0]\bar{E}
=[1.0,14.0]E_r,
\end{equation}
in order to investigate the out-of-equilibrium statistical relaxation dynamics in the quasiperiodic lattice.

\subsection*{C. Time Scales}%

The characteristic time scale of the system is determined by the energy unit $\bar{E}$ used in the MCTDHB calculations. Accordingly, the natural unit of time is defined as
\begin{equation}
\bar{\tau} \equiv \frac{\hbar}{\bar{E}},
\end{equation}
where $\bar{E}=\hbar^2/(m\bar{L}^2)=2E_r$. For the chosen lattice parameters with \(^{87}\mathrm{Rb}\) atoms, the recoil energy is \(E_r \approx 5.37\times10^{-30}~\mathrm{J}\), which yields the characteristic recoil time
\begin{equation}
\bar{\tau} \approx 9.82\times10^{-6}~\mathrm{s}
\approx 9.82~\mu\mathrm{s}.
\end{equation}

Throughout this work, all propagation times and dynamical observables are expressed in units of $\bar{\tau}$. The non-equilibrium dynamics are generated through the quasiperiodic lattice potential and subsequently evolved within the MCTDHB framework up to a maximum dimensionless propagation time of $t=1000$, corresponding to a physical evolution time of approximately $9.82~\mathrm{ms}$.

\section{Quantitative Characterization of Universal Hilbert-Space Relaxation}

To further quantify the emergence of universal relaxation dynamics in the strong-quench regime, we introduce a relative deviation measure that characterizes the approach of the Hilbert-space spreading measures toward their stationary values. For a generic quantity $X(t)$, chosen from the coefficient-based measures $S_C(t)$, $S_2(t)$, and $S_C^N(t)$, we define

\begin{equation}
\delta X(t)=
\frac{|X(t)-\langle X\rangle|}
{\langle X\rangle},
\end{equation}

where $\langle X\rangle$ denotes the long-time average of the corresponding quantity. The relative deviation $\delta X(t)$ measures the instantaneous distance from the stationary statistical state and therefore provides a direct characterization of the relaxation dynamics. 

\begin{figure}[t]
    \centering
    \includegraphics[width=0.6\columnwidth]{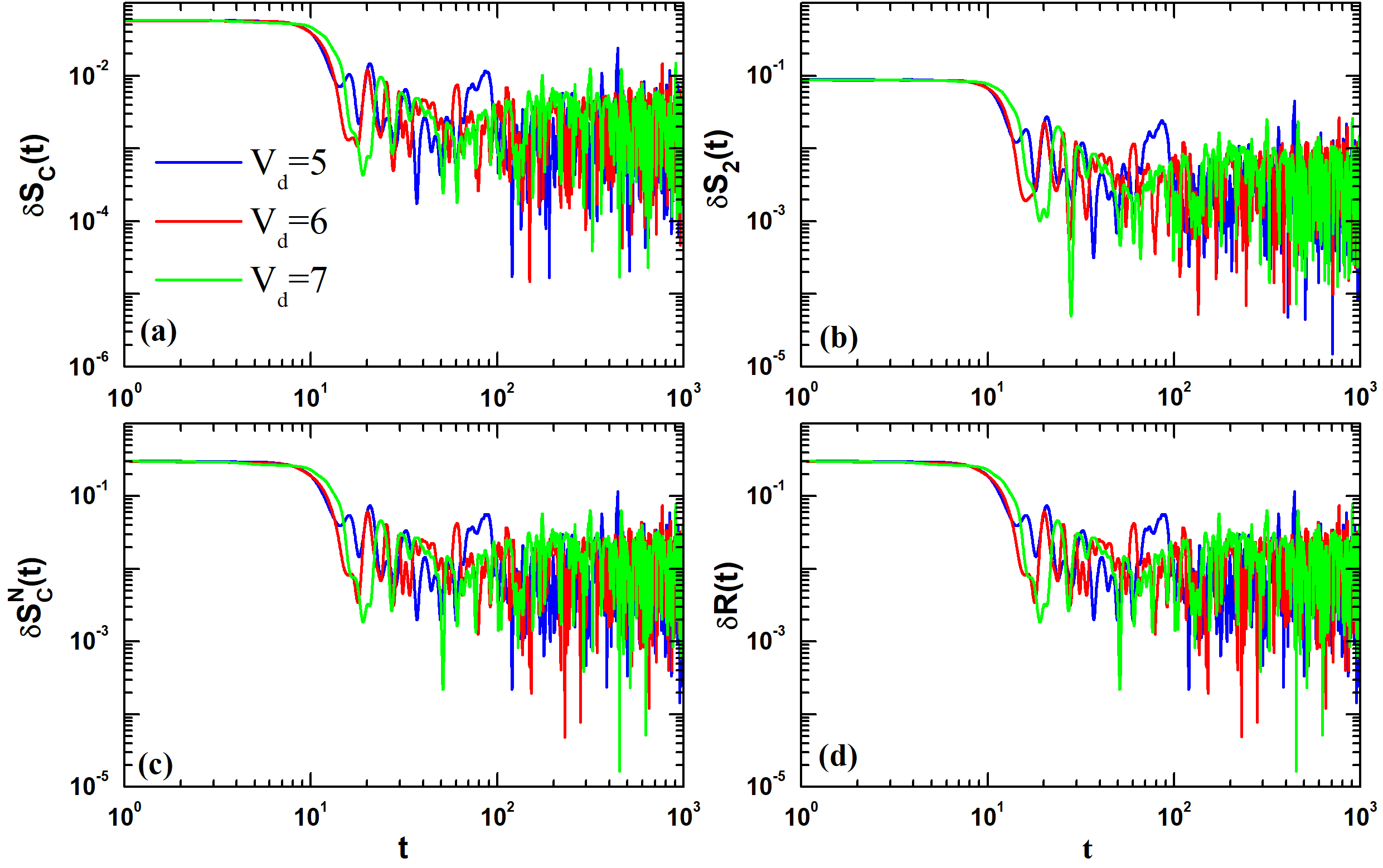}
\caption{Relative deviations of the Hilbert-space spreading measures from their long-time averaged values for strong quenches with $V_d=5,\,6,\,7$. (a) $\delta{S_C}(t)$, (b) $\delta{S_2}(t)$, (c) $\delta {S_C^N}(t)$, and (d) the combined relative deviation $\delta R(t)$. The collapse of the curves for different quench strengths supports, within the sampled strong-quench window, the presence of a common stationary relaxation trajectory in Hilbert space.  Parameters are $N=S=9$, $V_p=10$, $g_0=0.5$.}
    \label{fig3-deviation}
\end{figure}

Fig.~\ref{fig3-deviation} shows the relative deviations for the strong-quench regime with $V_d=5,6,7$. The individual deviations $\delta {S_C}(t)$, $\delta {S_2}(t)$, and $\delta {S_C^N}(t)$ exhibit a rapid decrease after the initial relaxation period, followed by bounded fluctuations around small stationary values. Remarkably, the curves corresponding to different quench amplitudes collapse onto a common dynamical profile within numerical accuracy. This collapse shows that, for the sampled values $V_d = 5, 6, 7$, the late-time dynamics become largely independent of the microscopic details of the perturbation and are governed by a common empirical relaxation process in Hilbert space.

To characterize this universality using a single quantity, we define a combined relative deviation incorporating all three Hilbert-space measures,

\begin{equation}
\delta R(t)=
\frac{
|S_C(t)-\langle S_C\rangle|
+
|S_2(t)-\langle S_2\rangle|
+
|S_C^N(t)-\langle S_C^N\rangle|
}
{
\langle S_C\rangle+\langle S_2\rangle+\langle S_C^N\rangle
}.
\end{equation}

The combined deviation exhibits the same universal collapse for $V_d=5,6,7$, confirming that the observed relaxation behavior is not a consequence of a particular definition but represents a robust feature, within the sampled quenches of the many-body wavefunction spreading. The residual fluctuations around the stationary values originate from coherent finite-size dynamics in the isolated finite system, while the overall decay of $\delta R(t)$ indicates approach to a common stationary coefficient regime for the sampled quenches. These results provide additional evidence for hidden GOE-like coefficient relaxation, where extensive many-body complexity develops despite the persistence of nearly unchanged low-order observables.

\section{Finite-Size Analysis}

We now briefly discuss the dependence of our results on system size to test qualitative robustness across the accessible sizes. While the main text focuses on \(N=9\) bosons in \(S=9\) lattice sites with \(M=10\) orbitals, we additionally perform simulations for smaller and larger systems within the same MCTDHB framework.

Specifically, we consider a smaller system with \(N=7\) bosons in \(S=7\) lattice sites using \(M=8\) orbitals, as well as a larger system with \(N=11\) bosons in \(S=11\) lattice sites using \(M=12\) orbitals. In all cases, the number of orbitals is chosen to ensure convergence of the many-body dynamics within the accessible computational resources.

All parameters of the quasiperiodic lattice, interaction strength, and quench protocol are kept identical to those used in the main text, allowing for a direct comparison of dynamical observables across different system sizes. This enables us to assess the stability of the observed statistical relaxation behavior against finite-size effects.

\begin{figure}[t]
    \centering
    \includegraphics[width=0.3\columnwidth]{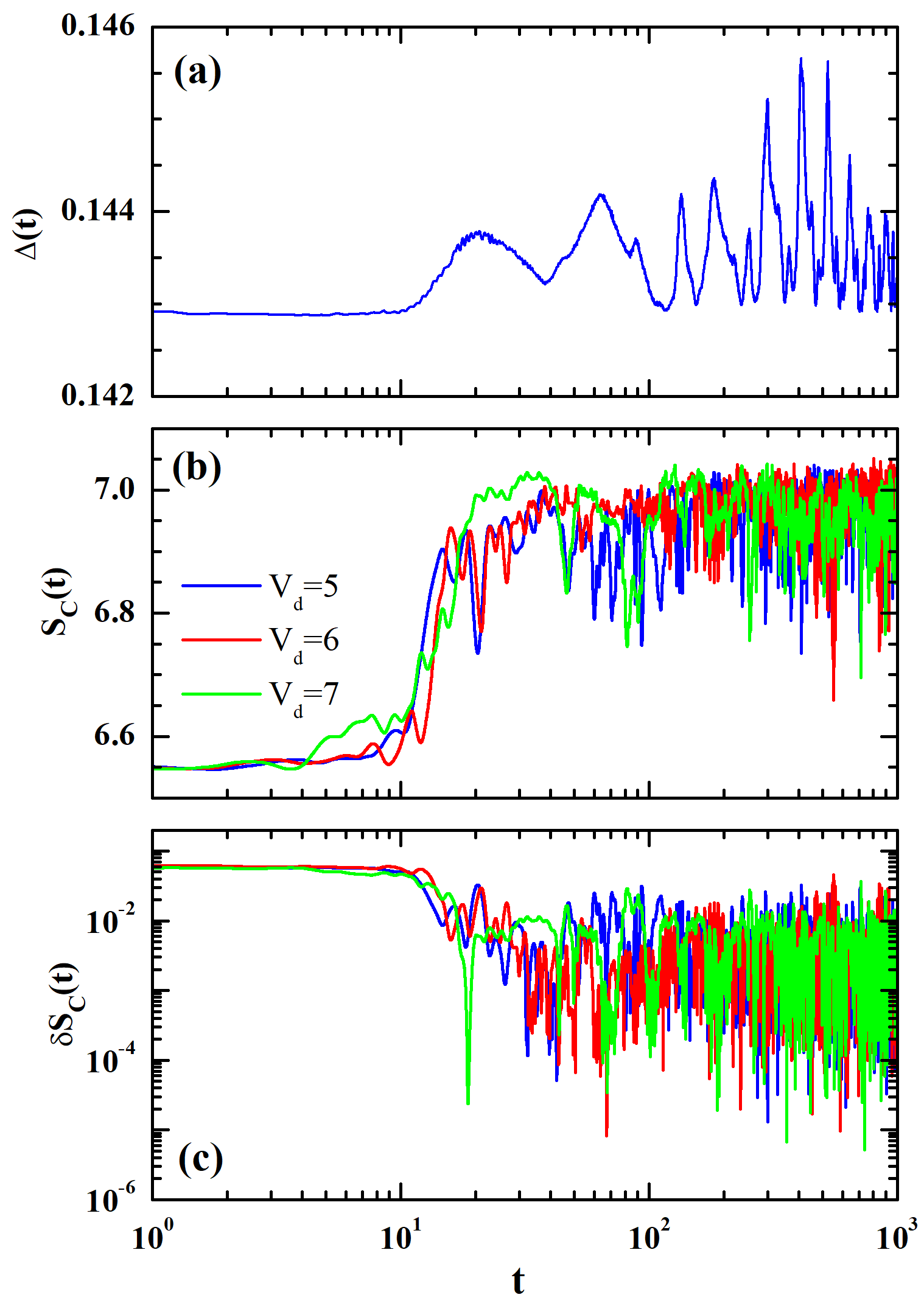}
\caption{Time evolution of the many-body observables for the smaller system size with $N=7$ bosons in $S=7$ lattice sites. (a) Order parameter $\Delta(t)$, showing fluctuations around the Mott-like value $1/S = 0.1428$, indicating persistence of the underlying correlated insulating structure. (b) Shannon entropy $S_C(t)$, exhibiting a three-stage relaxation dynamics characterized by an initial quadratic growth, intermediate linear increase, and eventual saturation. The long-time saturation value $S_C^{\mathrm{sat}} \approx 6.85$ is consistent with the Gaussian Orthogonal Ensemble (GOE) prediction for the corresponding finite Hilbert space. (c) Fluctuation of the Shannon entropy $\delta S_C(t)$, displaying the same qualitative relaxation behavior as $S_C(t)$, with saturation at long times. All quantities are plotted on a logarithmic time scale.
} 
    \label{fig-N7}
\end{figure}

For the smaller system size with \(N=7\) bosons in \(S=7\) lattice sites, we present representative results in Fig.~\ref{fig-N7} to verify the robustness of the statistical relaxation scenario. We analyze the order parameter \(\Delta(t)\), Shannon entropy \(S_C(t)\), and its fluctuation \(\delta S_C(t)\), all shown on a logarithmic time scale to resolve the different dynamical regimes.

The order parameter \(\Delta(t)\), shown in Fig.~\ref{fig-N7}(a), exhibits small fluctuations around the expected Mott-like value \(1/(S=7) = 0.1428\). Importantly, these fluctuations do not indicate any breakdown of the underlying Mott-type correlation structure, but rather confirm that the correlated insulating character remains intact throughout the dynamics, consistent with the discussion in the main text.

In contrast, the entropy \(S_C(t)\) [Fig.~\ref{fig-N7}(b)] displays a clear three-stage relaxation dynamics. At short times, a quadratic growth is observed, followed by an intermediate linear-in-time regime, and finally saturation at long times. With $N=7$ bosons and an effective orbital number $M=7$, the accessible many-body Hilbert-space dimension is $D=\binom{N+M_{eff}-1}{N}=1716$. Accordingly, the GOE estimate for the coefficient entropy is $S_C^{\mathrm{GOE}}\approx \ln(0.48D)=\ln(823.68)\approx 6.71$. The saturated value is found to be \(S_C^{\mathrm{sat}} \approx 6.85\), which agrees well with the Gaussian Orthogonal Ensemble (GOE) prediction for the corresponding finite Hilbert space of the \(N=7\), \(M=7\) system, indicating behavior consistent with the emergence of random-matrix-like coefficient behavior. Because the propagation used $M = 8$ adaptive orbitals whereas the nominal benchmark uses $M_{eff} = 7$, this agreement is indicative rather than a finite-size proof of GOE ergodicity.

Similarly, the fluctuation $\delta S_C(t)$ shown in Fig.~\ref{fig-N7}(c) exhibits the same qualitative features for the $N=9$ bosons in $S=9$ lattice sites system, including the emergence of a statistically stationary regime with bounded long-time fluctuations around a well-defined mean value. These results further support the robustness of the many-body relaxation dynamics against changes in system size, while the underlying Mott-like correlation structure remains preserved throughout the nonequilibrium evolution.

\begin{figure}[!tbh]
    \centering
    \includegraphics[width=0.3\columnwidth]{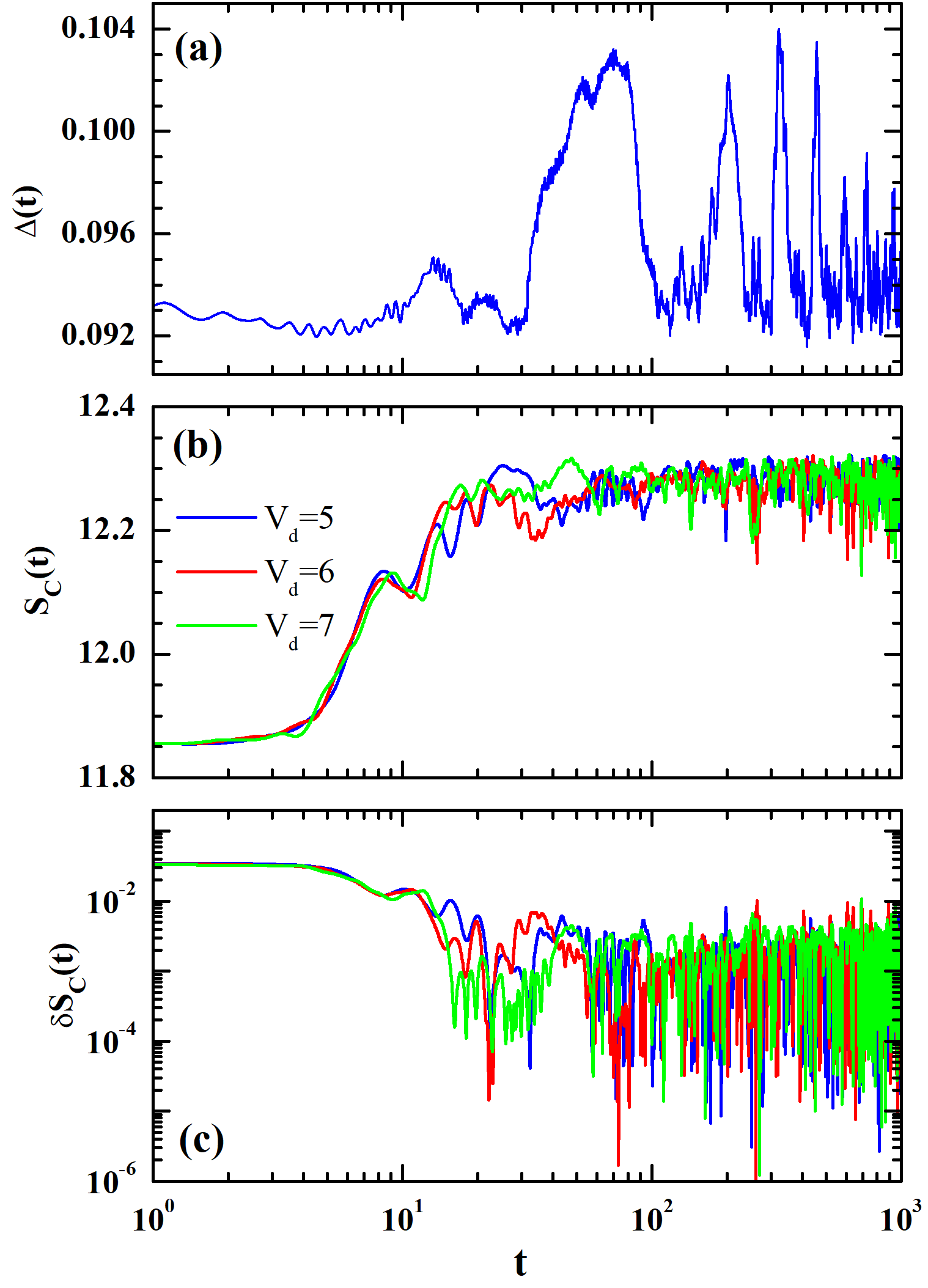}
\caption{
Time evolution of $\Delta(t)$, $S_C(t)$, and $\delta S_C(t)$ for the larger system size with $N=11$ bosons in $S=11$ lattice sites. (a) $\Delta(t)$ fluctuates and saturates around the Mott-like value $1/S=0.09$. (b) $S_C(t)$ shows a three-stage relaxation dynamics and approaches the GOE value at long times. (c) $\delta S_C(t)$ exhibits similar relaxation and saturation behavior. All quantities are shown on a logarithmic time scale.
}
    \label{fig-N8}
\end{figure}

For the larger system size with \(N=11\) bosons in \(S=11\) lattice sites, we observe qualitatively the same dynamical behavior as reported in the main text and in the smaller system analysis, confirming the robustness of the statistical relaxation scenario. The order parameter \(\Delta(t)\), shown in Fig.~\ref{fig-N8}(a), fluctuates and saturates around the expected Mott-like value \(1/(S=11) = 0.09\), reflecting the correct scaling of the underlying correlation structure with system size. As in the smaller system, this behavior indicates that the Mott-type correlations remain intact throughout the dynamics.

The Shannon entropy \(S_C(t)\) [Fig.~\ref{fig-N8}(b)] again exhibits a clear three-stage relaxation dynamics, consisting of an initial quadratic growth, followed by a linear increase in time, and eventual saturation at long times. The saturated value is consistent with the Gaussian Orthogonal Ensemble (GOE) prediction for the corresponding Hilbert space of the \(N=11\), \(M=11\) system, remaining consistent with random-matrix-like statistics in the long-time limit. Similarly, the fluctuation \(\delta S_C(t)\) shown in Fig.~\ref{fig-N8}(c) follows the same qualitative evolution, with growth at short times and saturation at long times. Overall, these results show qualitative persistence of the statistical-relaxation scenario over the sampled sizes; establishing finite-size convergence would require a systematic scaling analysis of effective-shell and spectral diagnostics.

\end{document}